\documentstyle[preprint,revtex]{aps}
\begin{document}
\draft
%\preprint{\today}
\begin{title}
Correlations for the S=1/2 Antiferromagnet on a Truncated Tetrahedron
\end{title}
\author{D.~Coffey$^a$ and S.~A.~Trugman$^b$}
\begin{instit}
$^a$ Center for Materials Science, MS-K765, Los Alamos National Laboratory,
Los Alamos, NM 87545
\end{instit}
\begin{instit}
$^b$ Theoretical Division, MS-B262, Los Alamos National Laboratory,
Los Alamos, NM 87545
\end{instit}
%\receipt{June, 1990}
\begin{abstract}

The antiferromagnetic Heisenberg Hamiltonian is investigated
on a truncated tetrahedron, which is a closed 12 site system.
We find that the
ground state has many similarities
to that of $C_{60}$.
We study 2- and 4-spin
correlations in the classical ground state of the
truncated tetrahedron and calculate the same correlations in the
exact S=1/2 ground state.  We find that the classical
correlations survive for a
range of bond strengths in the Heisenberg Hamiltonian
and that
one can construct a good trial wavefunction based on
the classical ground state.
This suggests that the correlations present in the classical ground state
of $C_{60}$ also survive in the exact ground state of that system,
for a
range of bond strengths about the physically relevant $J_2 \approx J_1$.
We calculate the momentum-space correlation function
$S ( q )$, which is measurable by neutron
scattering, for both $C_{12}$ and $C_{60}$.
We also calculate correlations at finite temperature.
\end{abstract}
\pacs{75.10-b, 75.25+z, 75.30.Kz and 75.30.-m}
\narrowtext

\section{Introduction}

We have found that an antiferromagnetic
Heisenberg Hamiltonian on $C_{60}$ has an
exotic ground state with nontrivial topology.\cite{ct}
The Heisenberg Hamiltonian
arises from a tight-binding Hamiltonian describing the half-filled
$\pi$ orbitals at the carbon sites, assuming that
there is strong
Coulomb repulsion between electrons in these orbitals.
The repulsion may be either long or short range.
The exotic classical $(S = \infty )$ ground state has
nontrivial topology, with a
skyrmion number 7.
(The vector field $\hat n$ that is normal
to the spin plane of the pentagons covers the sphere 7 times.)
$C_{60}$ is, however, a spin 1/2
system, and quantum corrections are expected to be important.
Estimates using perturbation theory indicate that the corrections
are about 60\% to the ground state energy and about 40\% to the
%are about 40\% to both the ground state energy and the
magnetization per site.
Given the size of these corrections, there is concern
that the perturbative results may be misleading.

The truncated tetrahedron $C_{12}$ is system
that is similar to $C_{60}$, but is small enough that the
$S= 1/2$ problem can be solved exactly.\cite{WCGK}
The classical ground state for $C_{12}$ also has nontrivial topology,
with a skyrmion number 1.
Here we
investigate 2- and 4-spin
correlations in the exact ground state of the $S=1/2$ Heisenberg
antiferromagnet on $C_{12}$, and compare to correlations in the classical
ground state.
We find that the classical correlations survive
in the exact ground state for a wide range of the
parameter $J_2 / J_1$, the ratio of
the two bond strengths in the system.
We also construct an $S= 1/2$ variational wavefunction
based on a coherent state representation of
the classical ground state spin configuration, and show that it
has a large overlap with the exact ground state.

These results suggest that the correlations
found in the classical ground state of
$C_{60}$ also survive for $S=1/2$ for a range of bond strengths.
There is expected to be
no phase transition in the exact ground state as a
function of on-site repulsion $U$ in the Hubbard model
on the $C_{60}$ lattice.\cite{ct,nopt}
The spin correlations for the large $U$ limit Heisenberg antiferromagnet
are therefore expected
to be present, although somewhat reduced in magnitude,
for the physically relevant case of intermediate $U$.
%Residual effects of these correlations in the doped $C_{60}$ remain to be
%investigated.

\section{Classical Approximation}

As in the case of $C_{60}$, all carbon atoms on
$C_{12}$ are equivalent, but there are two
inequivalent types of bonds.
The magnetic exchange constant for two neighboring sites
on the same triangle is $J_1 = t_1^2 / U,$
where $t_1$ is the hopping matrix element and $U$ is the Hubbard interaction.
The constant connecting a site
on one triangle with a nearest neighbor on another triangle is
$J_2= t_2^2 / U$.
The Hamiltonian is

\begin{equation}
H  = {J_1} \sum _ {<j,k>} ^ t \vec \sigma _j \cdot \vec \sigma _k
+ {J_2} \sum _ {<j,k>} ^ {nt} \vec \sigma _j \cdot \vec \sigma _k ,
\label{ham}
\end{equation}
where the first sum is over the 12 triangle bonds,
and the second over the 6 non-triangle bonds.
We take $J_1$ =1 and investigate the properties of this Heisenberg Hamiltonian
on $C_{12}$ as a function of $J_2$.
We first treat the Hamiltonian classically, so that $\vec \sigma$
in Eq. (\ref{ham})
is a classical unit vector.
This is the same as the $S \rightarrow \infty$
limit.  We discuss the properties of the exact
quantum $(S=1/2)$ ground state in the next section.

The truncated tetrahedron has 4 triangular faces
and 4 hexagonal faces.  An antiferromagnet is frustrated
on the triangular faces.  For an isolated triangle,
the ground state energy is $J_1 \cos ( 2 \pi / 3 ) = - J_1 / 2$ per bond,
which is smaller than the $- J_1$ per bond that an unfrustrated
system would have.  If the 12 spins on the truncated
tetrahedron are able to find an arrangement so that the
different triangles do not further frustrate each other,
the ground state energy will reach the lower bound
$E_b = 12 J_1 \cos ( 2 \pi / 3 ) - 6 J_2$.
The lower bound is an energy $- J_1 / 2$ per triangle bond,
and $-J_2$ per non-triangle bond.
The classical ground state configuration was found numerically by minimizing
the energy over the spin variables $\{ \theta _ i , \phi _ i \},~i=1,12$
(see Fig. 1).
We find that $C_{12}$ does attain the lower bound $E_b$, but
only in a topologically nontrivial spin arrangement.
In comparison,
$C_{60}$ also attains its lower bound, but other fullerenes
$C_{20}$, $C_{70}$, and $C_{84}$ do not.\cite{ct}

The ground state configuration
for $C_{12}$
is the same for all positive $J_1$ and $J_2$.
In addition to the global rotational symmetry of the ground state,
there is a discrete parity symmetry whereby each spin
$\vec \sigma _i  \rightarrow -\vec \sigma _i$.
The state related to a ground state $G_1$ by parity cannot be
reached by a global rotation on $G_1$.
Although the
ground state spin
configuration is non co-planar, it may be easily visualized.
There is a global rotation such that the spins
on each triangle lie in the plane of that triangle.
The skyrmion number $s$ for a set of vectors $\vec v _ i$ defined at points
$i$ on the surface of a sphere is given by
\begin{equation}
s = {1 \over {4 \pi}} \sum \alpha ( \vec v_i , \vec v_j, \vec v_k ),
\label{skyrm}
\end{equation}
where $\alpha $ is the solid angle subtended by three vectors and
the sum is over a set of fundamental triangles that cover
the sphere.
The skyrmion number for $C_{12}$ and $C_{60}$ is calculated by introducing
``twist'' bond variables $\hat n _i$
on the triangle bonds for the vectors $\vec v _ i$, with
\begin{equation}
\hat n_i \sim \vec \sigma _j \times \vec \sigma _k ,
\label{n}
\end{equation}
%as in the case of the pentagon bonds on $C_{60}$,
and one finds that both of the parity-related
ground states of $C_{12}$ have skyrmion number 1.
(The skyrmion number cannot be defined directly on
the spin variables because the solid angle in Eq. (\ref{skyrm})
is undefined for many sets of 3 neighboring spins.)
The $\hat n_i$ defined by any pair of spins
on the same triangle are equal.
The spin configuration may be chosen so that the $\hat n_i$
point along the normals to the triangles as shown in Fig. (1).
The sum of all $\hat n_i$ is zero.
The net magnetic moment in the ground state is also zero.

In the presence of an applied magnetic field, $C_{12}$ has a
{\it second order} metamagnetic
transition at which $dM / dh$ is discontinuous.
The skyrmion number is 1 both above and below the transition.
In contrast, $C_{20}$ and
$C_{60}$, both of whose ground states have skyrmion
number 7, have a first order metamagnetic transition at which
$M$ is discontinuous.

\section{Spin 1/2}

For $S=1/2$, the
$\sigma$ matrices in Eq. (\ref{ham}) represent
2 x 2 Pauli spin matrices.  (Our convention is to use
$\sigma$ matrices, which are a factor of 2 larger than $s$ matrices.)
The Heisenberg Hamiltonian on the truncated tetrahedron was solved in each
$ S _ z$ sector separately.\cite{affleck}
The dimension of the $ S _ z =0$ sector
is 924 and that of the $ S _ z = 1$ sector is 792, so that
the total spin $S=0$ subspace has dimension
$924-792 = 132$.
The lowest energy eigenvalue is present in the $ S _ z =0$ sector
and absent from the $ S _ z = 1$ sector, so the ground state
is a singlet.
We calculate various correlation functions as a function
of $J_2 / J_1$.  In $C_{60}$, $J_2$ is expected to be nearly
equal to, but slightly larger than $J_1$, because the corresponding
non-pentagon bond is about 3\% shorter than the pentagon
bond.\cite{yannoni}
$C_{12}$ has not been synthesized, but
we consider $J_2 \approx J_1$ to be the region of
physical interest.

First we consider the quantum two-spin correlations
$g_{ij} = < \vec \sigma _i \cdot \vec \sigma _j >$.
There are five inequivalent $g_{ij}$'s on the truncated tetrahedron,
which are represented by $g_{12}$, $g_{14}$, $g_{15}$, $g_{18}$ and $g_{19}$,
where the subscripts refer to the site labeling in Fig. (1).
The $g_{ij}$'s, with the exception of $g_{19}$, are plotted in Fig. (2).
(The classical $g_{19}$ is zero, and the quantum $|g_{19}| < 0.1$
for all $J_2$.)
The functions
$g_{12}$ and $g_{14}$ are the triangle and non-triangle bond energies
divided by the bond strengths.
One sees that as $J_2 \rightarrow 0$,
$g_{12}$ approaches -1, the value on an isolated triangle.
Surprisingly, as $J_2 \rightarrow 0$, the correlation
function $g_{14}$
remains large and negative.
When $J_2 \equiv 0$,
there is no correlation between spins on different triangles.
Then $< \vec \sigma_1 \cdot \vec \sigma_4 >$
is averaged over a manifold of degenerate ground
states and is equal to zero.  For $J_2$ small, however,
the expectation is over the unique ground state and is nonzero.
In the opposite limit of large $J_2$, $g_{14}$ approaches
-3 and all other $g_{ij}$ fall to zero.
This reflects the fact that in this limit,
the wavefunction becomes a product of singlet valence bonds
on the 6 non-triangle bonds, denoted $|VB>$.
The values of $g_{15}$ and $g_{18}$, the third and fourth nearest neighbor
correlation functions, both peak at $J_1 \approx J_2$.

We now calculate correlations between the spin plane normals
$\hat n_i$ in the exact ground state.  This will involve
four-spin correlation functions.
As we pointed out in the discussion
of the classical ground state,
the sum of the spin plane normals
$\sum _{i=1} ^4  \hat n_i$ for the four triangles is zero,
so that classically
\begin{equation}
\hat n _1 \cdot ( \hat n _2 +\hat n_3 + \hat n_4) = -1.
\label{nsum}
\end{equation}
%If the  $\hat n$ are not normalized the -1 on the right of Eq.(3) is changed
%to -0.75.
These correlations may be measured by

\begin{equation}
G = < \vec n_1 \cdot ( \vec n_2 + \vec n _3 + \vec n _4 ) >,
\label{G}
\end{equation}
where
\begin{equation}
\vec n _\alpha = (\vec \sigma _i \times \vec \sigma _j
 + \vec \sigma _j \times \vec \sigma _k
 + \vec \sigma _k \times \vec \sigma _i )/3.
\label{nvec}
\end {equation}
The spins on triangle $ \alpha$ are numbered counterclockwise
$(i,j,k)$.
(The $\hat n _\alpha$ in Eq. (\ref{nsum}) are normalized
to unit length, but the
$\vec n _\alpha$ in Eqs. (\ref{G}-\ref{nvec}) are not.)

The function $G$ measures the relative orientation of the spin-plane
normals, and involves
correlations between spins over the whole system.  It is plotted in Fig. (3).
As $J_2$ increases from zero, $|G|$ increases slowly
to its maximum at $J_2 / J_1 \approx 0.3$.
For large $J_2$, $|G|$ approaches zero as the wavefunction
approaches $|VB>$ and all correlations
other than those for the singlet
on nearest neighbor non-triangle bonds vanish.
The four-point function $G$ falls off at somewhat smaller
values of $J_2$ than the two-point functions in Fig. (2).
The quantum correlation functions will be
compared with classical and other correlation functions below.
Note that the skyrmion number itself is not a good
quantum number for the quantum system.

Singh and Narayanan\cite{SN} have emphasised that in identifying
a nonvanishing
expectation value of an operator in the ground state
as evidence for magnetic order, it is important to show that
the expectation value is considerably larger at zero than at
infinite temperature.  Note in this regard that $G$ and all of the
$g_{ij}$ are traceless operators and thus have
zero expectation value at $T=\infty$.

%\section{Trial Wavefunctions}
%
A classical ground state has each spin pointing in a fixed
direction.  The most straightforward translation of
a classical ground state into a spin 1/2 wavefunction
is to have each of the quantum spins point in the same fixed
directions.  The wavefunction for the first spin
is the two-component spinor $|c_1>$ that is the ground state
of $H_1 = -\vec h_1 \cdot \vec \sigma _1$, where $\vec h_1$
points in the classical spin direction.  These are coherent states.
The full wavefunction is
\begin{equation}
|C> = |c_1> |c_2> \dots |c_{12}>.
\label{coherent}
\end{equation}
Since the classical ground state configuration is
independent of $J_2/J_1$, so is the
trial wavefunction $|C>$ (and other trial wavefunctions discussed
below).
The expectation of the spin correlation functions
in the trial wavefunction $|C>$ and
in the exact groundstate $|\psi _ {ex}>$ for $J_1=J_2$
are shown in Table 1.
The exact ground state has large 2-spin and 4-spin
correlations, even at large separations.
The correlation functions $g_{ij}$ $(i \ne j)$
and $G$ are identical in $|C>$ and in the classical ground state.
The state $|C>$
does reasonably well with all of
the correlation functions, including $G$.  The largest error is the
underestimate of $g_{14}$, which has some singlet
character in $|\psi _ {ex}>$, but can have only Ising
character in $|C>$.

Any global rotation of a classical ground state gives
another ground state.  Thus a better trial state would
be to sum $|C>$ over all global rotations, or equivalently
to project it onto the singlet subspace.  As mentioned
in Section II, the parity operation
$\vec \sigma _i \rightarrow -\vec \sigma _i$ results in
another set of ground states that cannot be reached by rotations.
We therefore define states
$|\psi_S^1>$ and $|\psi_S^2>$ as the (normalized) singlet
projections of the parity related coherent states.
An additional related state is found by taking a linear
combination of these,
\begin{equation}
|\psi_S> = a_1 |\psi_S^1> + a_2|\psi_S^2>,
\label{paritysum}
\end{equation}
where $a_1$ and $a_2$ are complex numbers
with $|a_1|^2 + |a_2|^2 = 1$.
%%%%%
We determine $a_1$ and $a_2$ in Eq. (\ref{paritysum})
by maximizing $|<\psi_S|\psi_{ex}>|^2$.  (Virtually identical results
are obtained by minimizing the energy.)
The amplitudes
$a_1$ and $a_2$ are found to have equal magnitudes and
a phase difference that is almost independent of $J_2 / J_1$.
In Fig. (4) we plot the norm of the overlap squared of these states
and of $|VB >$ with the exact ground state
$|\psi_{ex}>$.
One sees that $|<\psi_S|\psi_{ex}>|^2$
and $|<\psi_S^1|\psi_{ex}>|^2$ are both peaked near $J_1=J_2$,
and that $|\psi_S>$ is a good approximation to the ground
state when $J_2 \approx J_1$.
$|VB>$ becomes exact for $J_2 \gg J_1$, but is already
failing in the region of physical interest, $J_2 \approx J_1$.

The correlation functions of the trial states
$|VB >$, $|\psi_S^1>$, and $|\psi_S>$ are given in Table 1.
($|\psi_S^2>$ has the same correlation functions as $|\psi_S^1>$,
and is not shown.)
The state $|VB>$ gives no indication of either the
magnitude or sign of any spin correlation function other
than $g_{14}$.
As the variational wavefunction
is improved in the sequence
$|C> \rightarrow |\psi_S^1> \rightarrow |\psi_S>$,
the correlation functions generally approach those
of the exact ground state.
Convergence is best at short distances and worst for the
long distance $g_{18}$.
By examining Figs. (2-3) and Table 1, one sees that
the trial state $|\psi_S>$ gives a reasonable approximation to
all of the correlation functions not just for $J_2 \approx J_1$, but
also as $J_2 \rightarrow 0$.
The agreement is remarkably good considering that
zero-point spin-wave fluctuations have not been added
to any of the trial states.

We now compare the energy of trial states and that
obtained from perturbation theory to the exact ground
state energy.
High energy contributions from
non-singlet states are removed
by projecting $|C>$ onto the singlet subspace.
The energy of $|\psi_S>$ is thus lower
than that of $|C>$.  The energy of $|\psi_S>$ is
-8.803 $J_1$ -12.325 $J_2$,
which is an upper bound on the exact ground state energy.
For $J_1 = J_2 = 1$, this energy is -21.128,
which is to be compared with an exact ground state energy of
-22.804.
{}From perturbation theory, the
leading correction to the ground state energy is
\begin{equation}
\Delta E = {\sum _ j} ^ \prime {{|<j | H | 0 > | ^ 2}
\over {\epsilon _0 - \epsilon _j}},
\label{pt}
\end{equation}
where $|0> = |C>$ and the sum is over intermediate states
$|j>$ with two adjacent spin flips.
(The ground state is determined by
stability against single spin flips.)
For $C_{12}$ this is
\begin{equation}
\Delta E = -6 {{J _2^2} \over {J_1}} -
{{27} \over {2}} {{J_1^2} \over {( 2 J_2 + J_1)}}.
\label{dele}
\end {equation}
The perturbed ground state energy is
$E^{(2)} = E_{cl} + \Delta E,$ with
$E_{cl}=-6 (J_1 + J_2 )$.
This expression clearly overestimates
the correction to the classical ground state energy
when $J_2 \gg J_1$, but it does fairly well for $J_2 \approx J_1$.
For $J_1 = J_2 = 1$, $E^{(2)} = -22.500$,
which should be compared with the exact ground state energy of -22.804.

For $C_{60}$, we expect the trial state $|VB>$
to be less successful than for $C_{12}$
in competing with the other trial
states.  The reason is that
$C_{12}$ contains highly frustrated triangle bonds,
which can be sacrificed at relatively low cost
by the formation of singlets on
the non-triangle bonds.
A pentagon is less frustrated than a triangle, with
a classical energy per bond of -.809, as compared to a triangle's
-0.5 and an unfrustrated bond's -1.
It is therefore reaonable to expect that the correlations found in
the classical ground state of $C_{60}$,
which with skyrmion number 7 are more exotic than those in the
truncated tetrahedron,
should survive for a larger range of $J_2$ values.

Neutron scattering experiments can measure $S(\vec q)$,
the Fourier transform of the spin-spin correlation function.
We calculate $S(q)$, which is $S$ averaged over all relative
orientations of $\vec q$ and the molecule.  This would be appropriate
for a sample containing molecules of random orientations.
Neglecting form factors,
\begin{equation}
S(q)  =  \sum _ {i,j}  <\vec \sigma _i \cdot \vec \sigma _j>
 \sin(q r_{ij})/q r_{ij},
\label{sq}
\end{equation}
where $r_{ij}$ is the cartesian distance between
sites $i$ and $j$.  We assume that the triangle
and non-triangle bonds lengths are equal.
The calculated S(q) for $|\psi_{ex}>$, $|\psi_S>$, $|VB>$, and the coherent
state $|C>$
are shown in Fig. (5).
(S(q) for $|C>$ is not zero at q=0 because $|C>$ is not a singlet.)
If the triangle and non-triangle bond lengths differ by very little,
as is the case for the pentagon and non-pentagon bonds in $C_{60}$,
it is difficult to separate the contributions to S(q) from
$g_{12}$ and $g_{14}$.
The longer distance spin correlations, present in all
of the wavefunctions except $|VB>$, cause higher frequency
oscillations in $S(q)$.  This leads, for example, to a shoulder
at $q r_{12} / \pi \approx 2.7$, which is strongest in $|C>$ but still visible
in the exact ground state.

For $C_{60}$, we do not know $|\psi_{ex}>$
and show $S(q)$ for the coherent state $|C>$
obtained from the classical ground state, and for the state $|VB>$
with singlets on the non-pentagon bonds.
There is a great deal more structure in S(q) for $|C>$ than for $|VB>$.
These features will, however, be smoothed out somewhat for
the exact ground state wavefunction for $C_{60}$ as they were for $C_{12}$.

{\it Finite temperature:}
We have calculated the $C_{12}$ correlation functions $g_{ij}$
at finite temperature, using the exact eigenfunctions
with Boltzmann weight.
As expected, all $g_{ij}$ for distinct spins approach
zero as $T \rightarrow \infty$.  Two of the correlation functions,
however, are nonmonotonic.  The correlation
$g_{12}$ between spins on the same triangle increases in magnitude
by 6\% with increasing $T$ before decreasing.
The correlation $g_{19}$, which is quite small at zero temperature,
actually changes sign and increases in magnitude by a factor
of 16 before decreasing to zero.

The finite temperature properties can help to interpret
recent Quantum Monte Carlo studies on the $C_{60}$
Hubbard model with 60 electrons by Scalettar et al.\cite{scalletar}
They find that for $U/t=8$, the spin correlations on $C_{60}$
resemble those of the classical ground state at
short and intermediate distances, but
decay to near zero at the largest separations.
A crucial question is whether
this long distance
decay is a property of the quantum ground state,
or whether it is caused by the finite temperature of the simulation.
For $C_{12}$, we calculated the exact $S=1/2$ Heisenberg
partition function
for $J$ corresponding to $U/t=8$ and the lowest temperature $T_1=t/5$
reached.  The temperature $T_1=0.582 \Delta$, where $\Delta$
is the gap to the first excited state.
The first excited state is, however, 9-fold degenerate
counting spatial and spin degrees of freedom, and higher
excited states are fairly close.
At $T=T_1$,
the system has a probability 0.19 to be in the quantum
ground state, and the correlation
function $g_{18}$ is 48\% of its zero temperature
value.
$C_{60}$ should require a lower temperature than $C_{12}$
for the long distance spin correlations to reach their zero
temperature values.
{\it If} the $C_{60}$ ground state had spin correlations extending across
the molecule, one would expect the $C_{60}$ excitation energies,
which correspond to magnons at allowed nonzero wavevector, to be
smaller than those of $C_{12}$ by roughly a factor of two
($\approx \sqrt {12/60}$).
By this estimate, $C_{60}$ at a given temperature would have
similar properties to $C_{12}$ at twice that temperature.
At $T=2T_1$, $C_{12}$ has a probability
0.028 to be in the ground state,
and a long distance correlation function
that is reduced by
a factor of 6 from its zero temperature value.
This suggests that the decay of spin correlations at large
distances observed in the Quantum Monte Carlo studies may be
a finite temperature effect.

\section{Summary and Conclusion}

The classical Heisenberg
antiferromagnet on the truncated tetrahedron
has an exotic ground state with nontrivial topology
and nonzero spin correlations up to the longest distances
in the system.
To see whether quantum fluctuations for $S=1/2$ destroy these
correlations, we exactly diagonalized the $S=1/2$ problem.
We find that the classical ground state gives a very
good indication of the exact quantum 2-spin and 4-spin
correlation functions for $J_1 \approx J_2$.
The classical ground state is also used to generate
trial $S=1/2$ wavefunctions
with good overlap against
the ground state for $J_1 \approx J_2$.
Although the wavefunction $|VB>$ with singlet valence
bonds on the non-triangle bonds is the correct one for $J_1 \ll J_2$,
this state gives no inkling of the long range correlations that
develop for the $J_1 \approx J_2$ region of physical interest.

This work has implications for the magnetic properties of $C_{60}$, in
which an exotic classical ground state has been found.
The results for $C_{12}$ suggest that the $S=1/2$ quantum
fluctuations will not destroy the classical spin correlations in $C_{60}$.
The quantum fluctuations may play a smaller role in $C_{60}$,
because the spin arrangement is less frustrated than that of $C_{12}$.
There is also concern that spin correlations in $C_{60}$ can be destroyed
because $U/t \approx 4$ is not large enough.
We have argued that the exact quantum ground state of
the Hubbard model in $C_{60}$ is unlikely to have a phase transition
as a function of $U/t$, and thus that the large $U$
spin correlations, somewhat reduced in magnitude, should be
observable in real $C_{60}$.\cite{ct}
Evidence for these magnetic correlations may be found in inelastic
neutron scattering experiments.
It may also be possible to infer the spin correlations by
doing NMR on a $C_{60}$ molecule with two $^{12}C$ nuclei
replaced by $^{13}C$.
It remains to investigate how doping affects the spin correlations,
and to what extent spin correlations
affect the properties of doped $C_{60}$ crystals,
such as $K_3C_{60}$.\cite{hebard}

\bigskip

The authors would like to thank A.~Balatsky, E.~Dagotto,
S.~Doniach, D.~Guo,
M.~Inui, D.~Rokhsar, and J.~R.~Schrieffer for
useful conversations. DC acknowledges the support of the Center for Materials
Science through the Program in Correlated Electron Theory.
This work was supported by the US Department of
Energy.

%\newpage

%\newpage
\begin{table}
\setdec 000.0000
\caption{Correlation functions for the trial wavefunctions
%$|\psi_S^+>$ and $|\psi_S>$,
and for the exact ground state $|\psi_{ex}>$
with $J_2=J_1$.}
\begin{tabular}{lccccc}
 &$|VB>$&$|C>$&$|\psi_S^1>$&$|\psi_S>$&$|\psi_{ex}>$\\
\tableline
$g_{12}$& 0& -0.5& -0.7745& -0.7336& -0.7295\\
$g_{14}$&-3& -1.0& -1.5890& -2.0541& -2.3416\\
$g_{15}$& 0&  0.5&  0.4761&  0.5131&  0.4453\\
$g_{18}$& 0& -0.5& -0.7745& -0.7323& -0.4916\\
$g_{19}$& 0&    0& -0.1088& -0.0418&  0.0013\\
G       & 0&-0.75& -0.4879& -0.5548& -0.5947\\
\end{tabular}
\label{table1}
\end{table}

\figure{Spin configuration on a
truncated tetrahedron.  A spin pointing up out of the plane
has a larger head, and one pointing into the plane has a larger tail.
%The spins on each triangle are coplanar.
For the global rotation shown, all spins lie in
the physical plane of their triangle, which is the plane
of the paper for triangle (123).
The truncated tetrahedron has been flattened by stretching
the back hexagon.
The spin directions are shown unchanged.
\label{spin} }

\figure{Two-spin correlation functions
$ g_{ij}=< \vec \sigma _i \cdot \vec \sigma _j> $,
$g_{12}$(dot-dash line),  $g_{14}$(dash line),  $g_{15}$(full line),
and  $g_{18}$(dot line),
as a function of $J_2 / J_1$.
\label{bond energies} }

\figure{Four-spin correlation function $G$, defined in Eq. (\ref{G}),
as a function of $J_2 / J_1$.
\label{4pt} }

\figure{Overlap squared of the trial wavefunctions with the exact
ground state,
$|<\psi_S|\psi_{ex}>|^2$ (full line),
$|<\psi_S^1|\psi_{ex}>|^2=|<\psi_S^2|\psi_{ex}>|^2$
(dot-dash line),
and $|<VB|\psi_{ex}>|^2$ (dot line).
\label {overlap}}

\figure{The angle-averaged spin correlation function $S(q)$
for $C_{12}$ is plotted as a function of $q r_{12} / \pi$
for $|\psi_{ex}>$(full line), $|\psi_S>$(dot line),
$|VB>$(dot-dash line), and $|C>$(dash line).
\label {sqtet}}

\figure{The angle-averaged spin correlation function $S(q)$
for $C_{60}$ is plotted as a function of $q r_{12} / \pi$
for $|C>$(full line) and $|VB>$(dash line).
\label {sqc60} }

\begin{references}

\bibitem[1]{ct} D.~Coffey and S.~A.~Trugman, preprint LA-UR-91-3302.

\bibitem[2]{WCGK} The ground state energy of the Hubbard model
on the truncated tetrahedron as a function of doping has
been studied by
S.~R.~White, S.~Chakravarty, M.~P.~Gelfand, and
S.~A.~Kivelson, Phys. Rev. B {\bf 45}, 5062 (1992).

\bibitem[3]{nopt} For a finite quantum system at zero
temperature, a phase transition occurs
as a function of the parameters only if the quantum
numbers change.  For $C_{12}$ and $C_{60}$, the only quantum
numbers are the spin $S$ and the angular momentum $L$
(actually what is left of $L$ under the finite rotation
group of the tetrahedron or icosahedron).  For both
$C_{12}$ and $C_{60}$, $S=L=0$ for $U/t=0$.
For $C_{12}$, we have determined from the exact
spectrum that $S=L=0$ in the Heisenberg limit $(U/t \rightarrow \infty)$.
For $C_{60}$, in the large $U$ limit, we know only that
the classical ground state has $\vec S$=0.
The simplest hypothesis consistent with these facts is that
the Hubbard model
has no phase transition as a function of $U/t$
for either $C_{12}$ or $C_{60}$.

\bibitem[4]{affleck} There is a more efficient valence bond
basis for calculating $S=0$ eigenfunctions, but this system
is small enough that we did not
need to use that method.
See K.~Chang, I.~Affleck, G.~W.~Hayden, and Z.~G.~Soos,
J.~Phys.~Condens.~Matter {\bf 1}, 153 (1989).

\bibitem[5]{yannoni} C.~S.~Yannoni et al., J.~Am.~Chem.~Soc.
{\bf 113}, 3190 (1991).

\bibitem[6]{SN} R. R. Singh and R. Narayanan,
Phys. Rev. Lett. {\bf 65}, 1072 (1990).

\bibitem[7]{scalletar} R.~T.~Scalettar, E.~Dagotto,
L.~Bergomi, T.~Jolicoeur, and H.~Monien, Florida State University preprint.

\bibitem[8]{hebard} A.~F.~Hebard et al., Nature {\bf 350}, 600 (1991).

\end{references}
\end{document}